\pdfoutput=1
\RequirePackage{ifpdf}

\documentclass{JINST}

\usepackage{graphicx}

\usepackage[
  backend=bibtex8,
  maxnames=6,                         
  bibstyle=numeric,               
  hyperref=true,                  
  sorting=none
]{biblatex}
\bibliography{SMFIPaper}

\title{Single Molecule Fluorescence Imaging as a Technique for Barium Tagging in Neutrinoless Double Beta Decay}

\author{B. J. P. Jones, A. D. McDonald and D. R. Nygren\\
\llap University of Texas at Arlington,\\
  502 Yates St, Arlington, TX 76019, United States of America\\
  E-mail: \email{ben.jones@uta.edu}, \email{austin.mcdonald@uta.edu}, \email{nygren@uta.edu}}

\abstract{Background rejection is key to success for future neutrinoless double beta decay experiments. To achieve sensitivity to effective Majorana lifetimes of $\sim10^{28}$ years, backgrounds must be controlled to better than 0.1 count per ton per year, beyond the reach of any present technology.   In this paper we propose a new method to identify the birth of the barium daughter ion in the neutrinoless double beta decay of $^{136}$Xe.  The method adapts Single Molecule Fluorescent Imaging, a technique from biochemistry research with demonstrated single ion sensitivity.  We explore possible SMFI dyes suitable for the problem of barium ion detection in high pressure xenon gas, and develop a fiber-coupled  sensing system with which we can detect the presence of bulk Ba$^{++}$ ions remotely.  We show that our sensor produces signal-to-background ratios as high as 85 in response to Ba$^{++}$ ions when operated in aqueous solution.  We then describe the next stage of this R\&D program, which will be to demonstrate chelation and fluorescence in xenon gas.  If a successful barium ion tag can be developed using SMFI adapted for high pressure xenon gas detectors, the first essentially zero background, ton-scale neutrinoless double beta decay technology could be realized.}

\keywords{Gaseous detectors;Scintillators, scintillation and light emission processes (solid, gas and liquid scintillators); Very low-energy charged particle detectors; neutrinoless double beta decay}

\begin{document}

\section{Signals and backgrounds in neutrinoless double beta decay}

The nature of the neutrino is one of the open fundamental questions of nuclear and particle physics.  The existence of a non-zero neutrino mass has been established through the discovery of flavor oscillations. This allows for the possibility that neutrinos are Majorana fermions, with deep implications for cosmology and particle physics.  The only known experimental method to establish the Majorana nature of the neutrino is a robust observation of neutrinoless double beta decay ($0\nu\beta\beta$). 

The process of double beta decay is possible in even-even nuclei in which the nucleus $^A_Z$X is stable relative to  $^A_{Z+1}$X but unstable relative to  $^A_{Z+2}$X.  In these cases, the nucleus can, according to the standard model, decay via the reaction $^A_{Z}X \rightarrow ^A_{Z+2}X + 2e^- + 2\bar{\nu}$, with the final state kinetic energy shared between the two detectable electrons and two invisible neutrinos.  If and only if the neutrino is a Majorana fermion, a similar process with no neutrinos in the final state may occur,  $^A_{Z}X \rightarrow ^A_{Z+2}X + 2e^-$. 

Various isotopes have been used as target media in $0\nu\beta\beta$ searches.  The strongest limits on lifetime, which under most nuclear matrix element models also correspond to the strongest effective Majorana mass limits, have been obtained using $^{136}$Xe \cite{KamLAND-Zen:2016pfg,Albert:2014awa}.  The present limit sits at $T_{0\nu\beta\beta} > 1.07\times10^{26}$ yr at 90\% CL  \cite{KamLAND-Zen:2016pfg} .  Discovery of $0\nu\beta\beta$ is possible at any lifetime beyond this limit if the neutrino mass ordering is normal, or in a range of lifetimes reaching up to $\tau\sim10^{28}$ yr if the ordering is inverted.  If the mass ordering is normal, discovery of $0\nu\beta\beta$ would also represent a measurement of the absolute neutrino mass scale.  Given the inverted ordering, on the other hand, the absolute mass scale can likely not be determined, but the Majorana nature is falsifiable as well as discoverable.  A goal for future $0\nu\beta\beta$ searches is to reach sensitivities near the maximum lifetime in the inverted scenario \cite{NSACLongRange} which is around a hundred times beyond present experimental limits.  Reaching this sensitivity with a ton scale detector in a practical scenario of a few years of running requires backgrounds to be controlled at the level $b$ $\lesssim$ 0.1, with background index $b$ expressed in units of counts per ton per year per in the energy region of interest (ROI).  This condition is, unfortunately, not satisfied by any existing technology.

The backgrounds to $0\nu\beta\beta$ searches can be divided into two classes.  The first of these is the two-neutrino process $^A_{Z}X \rightarrow ^A_{Z+2}X + 2e^- + 2\bar{\nu}$, which is expected to occur at a rate at least $10^6$ higher than the neutrinoless mode.  The only way to distinguish the neutrinoless mode from the two-neutrino mode is by sufficiently precise reconstruction of the event energy.  For xenon detectors, an energy resolution better than $\lesssim$2\%~FWHM at the Q-value for $0\nu\beta\beta$, $Q_{\beta\beta}$, is required to achieve a background index of $b_{2\nu}\lesssim$ 0.1 from the two-neutrino mode.  Energy resolution is sufficiently critical for the sensitivity of the experiment that a significant margin of safety in the experimental design is desirable.  An energy resolution of 1.0\% FWHM at 662 keV has been demonstrated using high pressure gaseous xenon (HPGXe) detectors \cite{Tpc2012}.  If extrapolation to $Q_{\beta\beta}$ is not compromised by systematics, an energy resolution of 0.57\% may be realized.  The presently commissioning NEXT-NEW detector will test the extrapolation of this resolution to larger experimental scales \cite{Collaboration2016}.  The energy resolutions of present generation liquid xenon (LXe) and liquid-scintillator-dissolved xenon (LSXe) detectors are 3.6\% and 11.1\% FWHM respectively \cite{KamLAND-Zen:2016pfg,Albert:2014awa}, giving background indices from the two-neutrino mode of 1 and 84  for these two technologies \cite{KamLAND-Zen:2016pfg,EXOICHEP}.

The second class of backgrounds is from ambient radioactivity and detector materials.  None of the contemporary experimental techniques that have placed significant limits has been free of such backgrounds.  Levels of experimentally determined or projected background for current contenders lie in the range 4 < $b$ < 300 according to an independent assessment \cite{LRP}.  That assessment reported overall background indices, after all cuts, of 9 (HPGXe), 130 (LXe) and 210 (LSXe)\footnote{Since the report, a campaign to remove $^{110m}$Ag from Kamland-Zen has led to reduction of this background.  Publication \cite{KamLAND-Zen:2016pfg}, reporting only the background in the inner 25\% of the fiducial volume, shows a factor two improvement.}, all significantly in excess of the target value $b\lesssim0.1$.  There are four known ways to further remove radioactive backgrounds:

\begin{itemize}
  \item Improve energy resolution
  \item Implement or improve event topology reconstruction
  \item Improve radiopurity and / or shielding
  \item Implement decay daughter identification (tagging)
\end{itemize}

Improved energy resolution allows a smaller ROI to be defined around  $Q_{\beta\beta}$, thus reducing contributions from backgrounds with broad energy spectra.  The energy resolutions of various xenon-based technologies were discussed above.  Event topological reconstruction involves distinguishing the two-electron signature of real double beta decay events from the signatures of beta- and gamma-induced backgrounds. In HPGXe detectors, for example, simulations indicate that a factor of approximately 10$^{-7}$ background suppression for $\gamma$-induced events can be obtained through differences in topological characteristics. 

Despite the clear power of energy and topological reconstruction as background rejection tools, achieving a further factor of 100 improvement for the present lowest-background technologies appears to be a formidable task.  The third method, background reduction by improved radio-purity and shielding, represents an attempt to assert that an event satisfying all true event criteria is unlikely to be background; credibility thus depends crucially on credibility of a background model.  At the ton-scale, it will be extremely difficult to demonstrate {\em{a priori}} that the background level is $b<0.1$, however heroic the radio-purity campaign. 

An alternative approach -- to assert that an event is both very likely to be a true event {\it and} very unlikely to be a background event -- may turn out to be an essential step.  Detection of creation of a daughter atom, in space and time, is the only known concept to realize such a powerful event-by-event criterion.  The double beta decay of $^{136}$Xe in either gas \cite{Elba} or liquid \cite{Brunner:2014sfa,Flatt:2007aa,Sinclair:2011zz}, appears to offer opportunities to develop this capability.  

Both HPGXe and LXe time projection chambers (TPCs) are being actively pursued as possible $0\nu\beta\beta$ technologies.  Each medium has distinct advantages and disadvantages: HPGXe provides fine energy resolution and topological signature, whereas LXe offers the possibility of self-shielding using a monolithic volume of enriched xenon to produce an effectively lower-background inner region.  The two media also invite different approaches to the problem of barium tagging.  To our knowledge, no barium tagging method has yet been proposed for LSXe detectors.

In this paper we present a novel barium tagging method for HPGXe detectors using single molecular fluorescent imaging (SMFI)  \cite{valeur2012molecular,lakowicz2013principles}.    SMFI is based on conversion of weakly fluorescent molecular precursors into strongly fluorescent ones by chelation with doubly charged ions, and has become a major element in the arsenals of contemporary biology and biochemistry\footnote{The 2014 Nobel Prize in Chemistry was awarded to three physicists for their seminal contributions to SMFI.}.  In xenon gas, at or near room temperature, it appears plausible that single barium ions may be captured leading to a detectable fluorescent state. If this is the case, these techniques may be  extendable to the problem of $0\nu\beta\beta$ at the ton-scale and allow the realization of an essentially background-free experiment.

\section{Xenon TPC detectors and barium tagging}

Detection of the daughter atom in $0\nu\beta\beta$  has been long
recognized as a strong positive criterion for discovery \cite{Moe:1991ik}, since
no conventional process can introduce a new atom with Z+2. In the decay $^{136}$Xe to barium, with
or without neutrinos, it is very likely that the disruptive departure of the nascent
electrons from the nucleus will leave the daughter barium atom in a highly ionized state \cite{PhysRev.107.1646}.  The barium ion rapidly captures
electrons from nearby neutral xenon atoms until
further capture is energetically disfavored. The process stops at the doubly ionized
state Ba$^{++}$ because the second ionization potential of barium (10.004
eV) is very far below the first ionization potential of xenon
(12.14 eV), relative to kT.  Depending on xenon density, nearby free electrons liberated by the two emergent beta particles can further neutralize the Ba$^{++}$ ion.  In LXe, electrons thermalize very near to sources of ionization leading to significant charge recombination, so the taggable barium ions  are expected to be distributed across the Ba, Ba$^{+}$ and Ba$^{++}$ charge states. This expectation is supported by measurements made by the EXO collaboration, which show a distribution of charge states in ions produced in decays of radon daughters in LXe  \cite{PhysRevC.92.045504}.  In pure xenon gas, at atomic densities nearly two orders of magnitude smaller than LXe, our simulations indicate that recombination is very unlikely and that the dication Ba$^{++}$ will be the dominant outcome.  This feature alone implies that different barium tagging methods may be optimal for each medium.

\subsection*{Previous barium tagging work in liquid xenon}

To detect the barium daughter in LXe, attention has been
given by others to spectroscopic features of the singly ionized state, Ba$^{+}$.
This atomic configuration permits a sequence of repetitive excitation/de-excitation cycles with alternating red and blue light involving a
long-lived triplet D state \cite{Moe:1991ik}. However, this
sequence must be performed in an ion trap, which must be held at a 
high quality vacuum to prevent collisional broadening \cite{Danilov:2000pp} and rapid quenching of the D state. Repeated detection
of an alternating sequence of photons of two colors from a single
atom is nevertheless a robust determination of the presence of Ba$^{+}$. 

An alternative approach is to perform the fluorescence cycle in a crystal of frozen xenon \cite{Mong:2014iya}.
To identify barium ions in a running LXe detector using this method, a cryogenic
probe must be inserted into the active volume of the TPC at the position
of an observed $0\nu\beta\beta$ candidate to execute an electrostatic capture. After electrostatic capture of the ion, a localized region is frozen. This solid xenon block can be either spectroscopically probed
{\em in situ}, or extracted for probing outside the detector \cite{Twelker:2014zsa}. In practice, the sequence
of extraction from a large mass of high-purity
cryogenic liquid, reliable conversion to a singly charged state, efficient
transport into a low-pressure trap or solid probe region and extended spectral interrogation, with credible spatial and temporal
correlation to the $0\nu\beta\beta$ event candidate, is clearly challenging and is not yet demonstrated.

\subsection*{Previous barium tagging work in gaseous xenon}
Gas-phase barium tagging has also been explored \cite{Brunner:2014sfa,Flatt:2007aa,Sinclair:2011zz}.
 Because Ba$^{++}$
has a noble-gas-like electronic shell configuration, it does not have
low-energy transitions which can be exploited for atomic fluorescence tagging.
For this reason, the primary approach to gas-phase barium tagging
has been to convert Ba$^{++}$ into Ba$^{+}$, transport
it efficiently to from its origin to a vacuum trap via an RF carpet and differential pumping, then capture and probe the singly-charged state using the two-color method.

\section{Single molecule fluorescent imaging}

\begin{figure}[b]
\begin{centering}
\includegraphics[width=0.90\columnwidth]{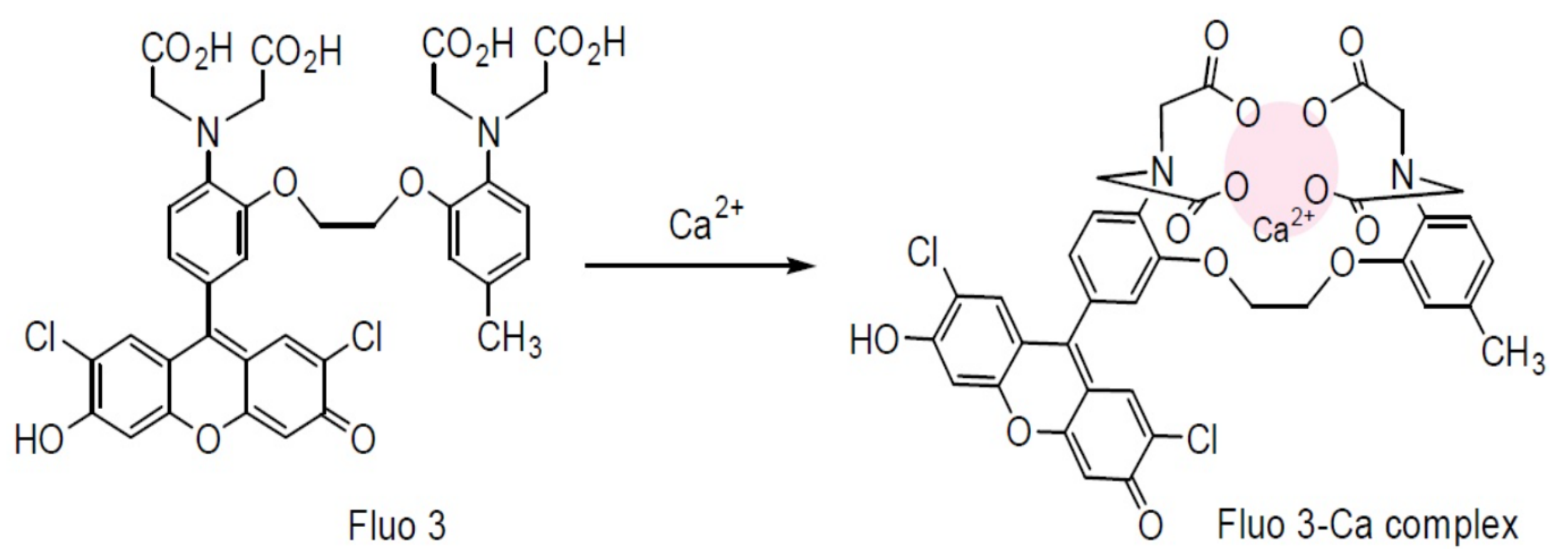}
\par\end{centering}

\caption{The structure of Fluo-3 is shown before and after complex formation \cite{Dojindo}.   \label{fig:fluo3}}
\end{figure}

In this paper we describe a new method for automatic {\em{ in situ}} identification of Ba$^{++}$ in high pressure xenon gas, based on the technique of Single Molecular Fluorescence Imaging. 
Developed originally by physicists, Single Molecular Fluorescence Imaging (SMFI) has been adopted
and advanced by biologists and chemists to an array of highly sophisticated techniques.  In SMFI, a small optically
thin region is interrogated repeatedly with typically blue or near-UV photons
that excite a molecule of interest. If the molecule is complexed with
a doubly charged ion, often Ca$^{++}$ in biological studies, the resulting adduct fluoresces strongly, whereas ion-free un-complexed molecules respond very weakly. Image-intensified
CCD cameras are used to detect single photons with pixel-scale spatial
resolution. Repeated interrogations provide statistically precise
identification and localization of a single molecule\textemdash even
inside living cells. The interrogation
rate can exceed 10$^5$ per second and fluorescence quantum yields
approach unity in many cases. Fluorescence detection
is facilitated by an inherent Stokes shift or a delay in response
time relative to pulsed excitation.

A wide variety of molecules for SMFI purposes now exists. Of interest are fluorophores such as Fluo-3, Fluo-4, and Rhod-2 which chelate
Ca$^{++}$. Calcium chelation by Fluo-3, one of the most commonly used calcium sensitive in fluoresence microscopy, is shown schematically in Fig.
\ref{fig:fluo3} \cite{Dojindo}. Fluo-4 is chemically similar to Fluo-3, but with the chlorine atoms replaced by fluorine for a higher light yield and increased stability against photobleaching.  Rhod-2 is a derivative of rhodamine with correspondingly longer wavelength fluorescence response, commonly used in two-color imaging.  For this work we have focussed on Fluo-3 and Fluo-4 dyes, though Rhod-2 and dyes more specially tailored for barium \cite{Nakahara2004} appear to hold equal promise given our current understanding.

\begin{figure}[t]
\begin{centering}
\includegraphics[width=0.99\columnwidth]{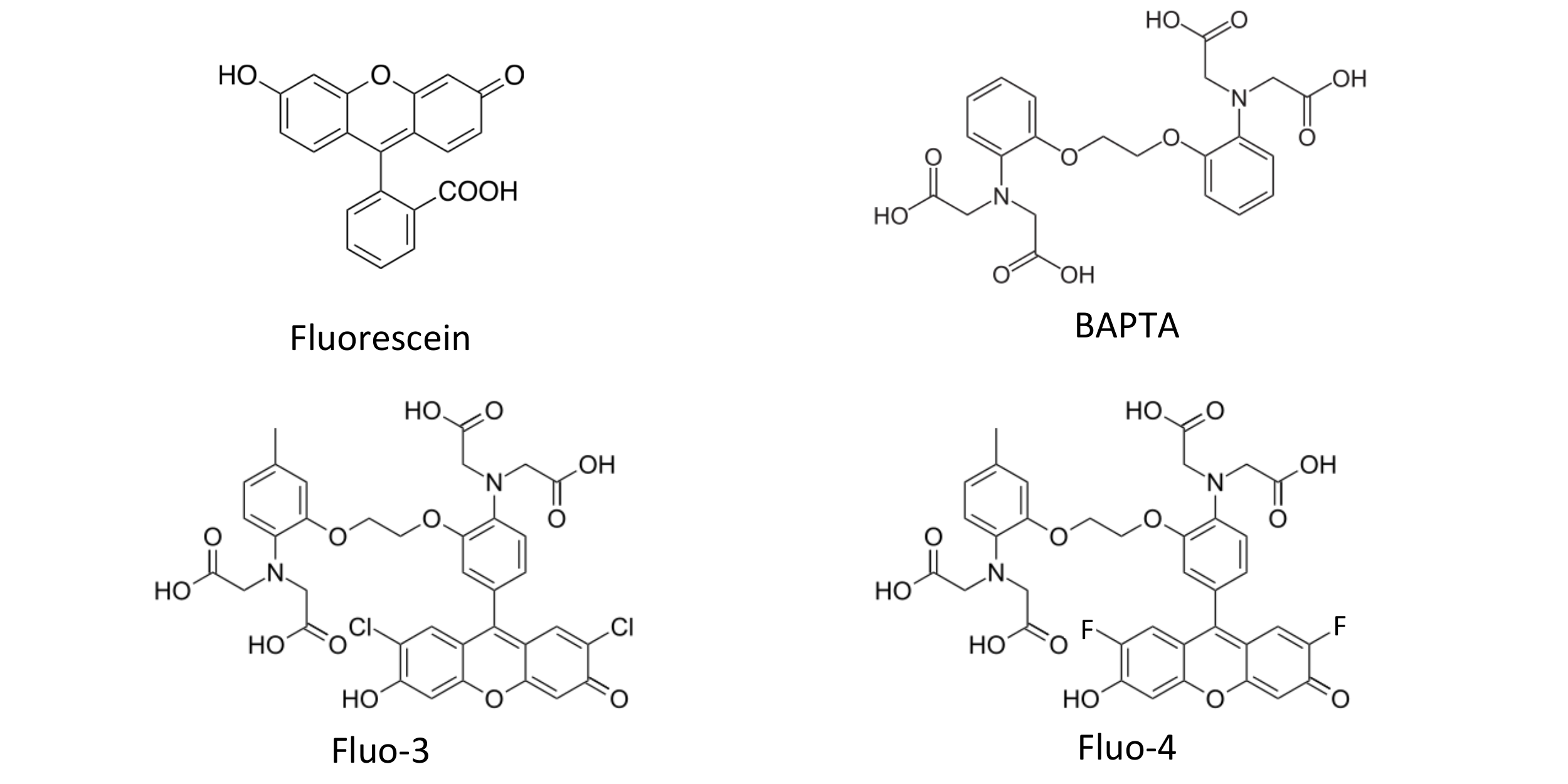}
\par\end{centering}

\caption{Structures of some important molecules described in this paper : fluorescein, BAPTA, Fluo-3 and Fluo-4. \label{fig:fluorescein}}
\end{figure}

\begin{figure}[t]
\begin{centering}
\includegraphics[width=0.99\columnwidth]{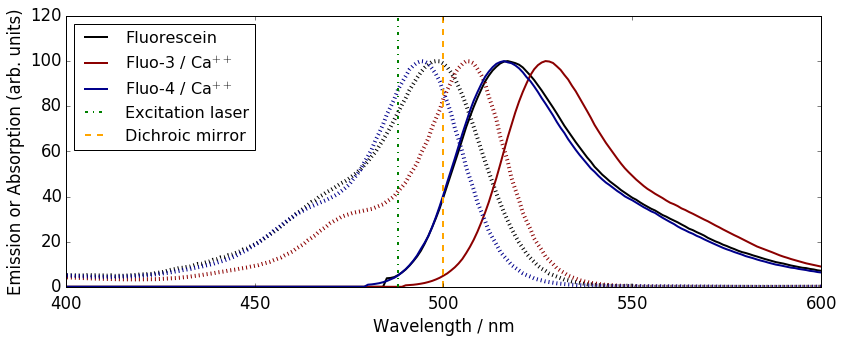}
\par\end{centering}

\caption{Absorption (dashed) and emission (solid) spectra of fluorescein precursor, and chelated Fluo-3 and Fluo-4 dyes, using data reported by \cite{ThermoSpectra} \label{fig:spectrafromthermo}}
\end{figure}

Fluo-3 and Fluo-4 are both derived from the fluorescein molecule, which is shown in Figure  \ref{fig:fluorescein}, top left.  The excitation and emission spectra of fluorescein, calcium chelated Fluo-3 and calcium chelated Fluo-4 are shown in Figure \ref{fig:spectrafromthermo}.  All these fluors have excitation that peaks in the blue with emission in the green, though Fluo-3 has a somewhat different wavelength dependence to the others, making it slightly less bright at our chosen excitation wavelength of 488 nm.  Fluo-3 and Fluo-4, shown in Figure \ref{fig:fluorescein}, bottom left and right, effectively consist of a fluorescein-like molecule bonded to a calcium-chelating molecule called BAPTA, which is shown in Figure \ref{fig:fluorescein}, top right.   When the molecule is isolated from doubly charged ions, the BAPTA-like part has many vibrational degrees of freedom,  allowing de-excitation of the fluorescein-like part without emission of a photon.  Thus non-chelated Fluo-3 and Fluo-4, when excited with blue light, de-excite non-radiatively. In the presence of Ca$^{++}$, on the other hand, the BAPTA-like part of the molecule forms a rigid cage around the ion, which has the effect of redistributing electrons within the complex and preventing vibrational de-excitation, thus restoring the fluorescence emission from the fluorescein-like part.  In this way, the dye transitions from a non-fluorescent state to a fluorescent one in the presence of Ca$^{++}$, making it a powerful tool for calcium sensing \cite{Thomas2000,Oliver2000,Paredes2009}.

Quoted response ratios between unchelated and calcium-chelated
states of Fluo-3 and Fluo-4 vary from 60 to more than 100 in biological milieu. Although
barium is uncommon in biochemistry research, the fact that barium
and calcium are congeners suggests that techniques for Ca$^{++}$ may have
relevance for Ba$^{++}$, and that dyes developed
for calcium sensitivity may be used directly for barium tagging.   In the next section
we study the properties of Ba$^{++}$-chelated Fluo-3 and Fluo-4, and implement a detection system
which we intend to use for studying dye chelation remotely inside a xenon gas environment.

\section{SMFI for barium detection \label{sec:LiquidTests}}

The difficult process of realizing SMFI-based barium tagging can be broadly broken into four steps of increasing difficulty: 1) identify dyes which provide a strong fluorescent response to barium dications;  2) develop a scanning system which can be used to tag barium ions remotely inside a large detector; 3) establish whether the chelation and fluorescence behavior is maintained in a HPGXe environment; 4) optimize the detection technique to the single molecule regime.  In this section we describe progress on items (1) and (2), and Section \ref{sec:GasTests} will briefly outline plans to address item (3) as the next immediate goal. Extension to the single molecule regime is not described in this paper, though single ion detection with these techniques has precedents, both in aqueous solutions and in living organisms.  Assuming chelation does occur in gas, single ion sensitivity in HPGXe is plausibly achievable, helped by the fact that the inert environment of a HPGXe detector is expected to be more favorable to survival of the dye/Ba$^{++}$ complex than aqueous environments where photobleaching by reactive oxygen species is a limiting effect \cite{Thomas2000,Oliver2000}.

\begin{figure}[t]
    \centering
    \includegraphics[width=0.99\textwidth]{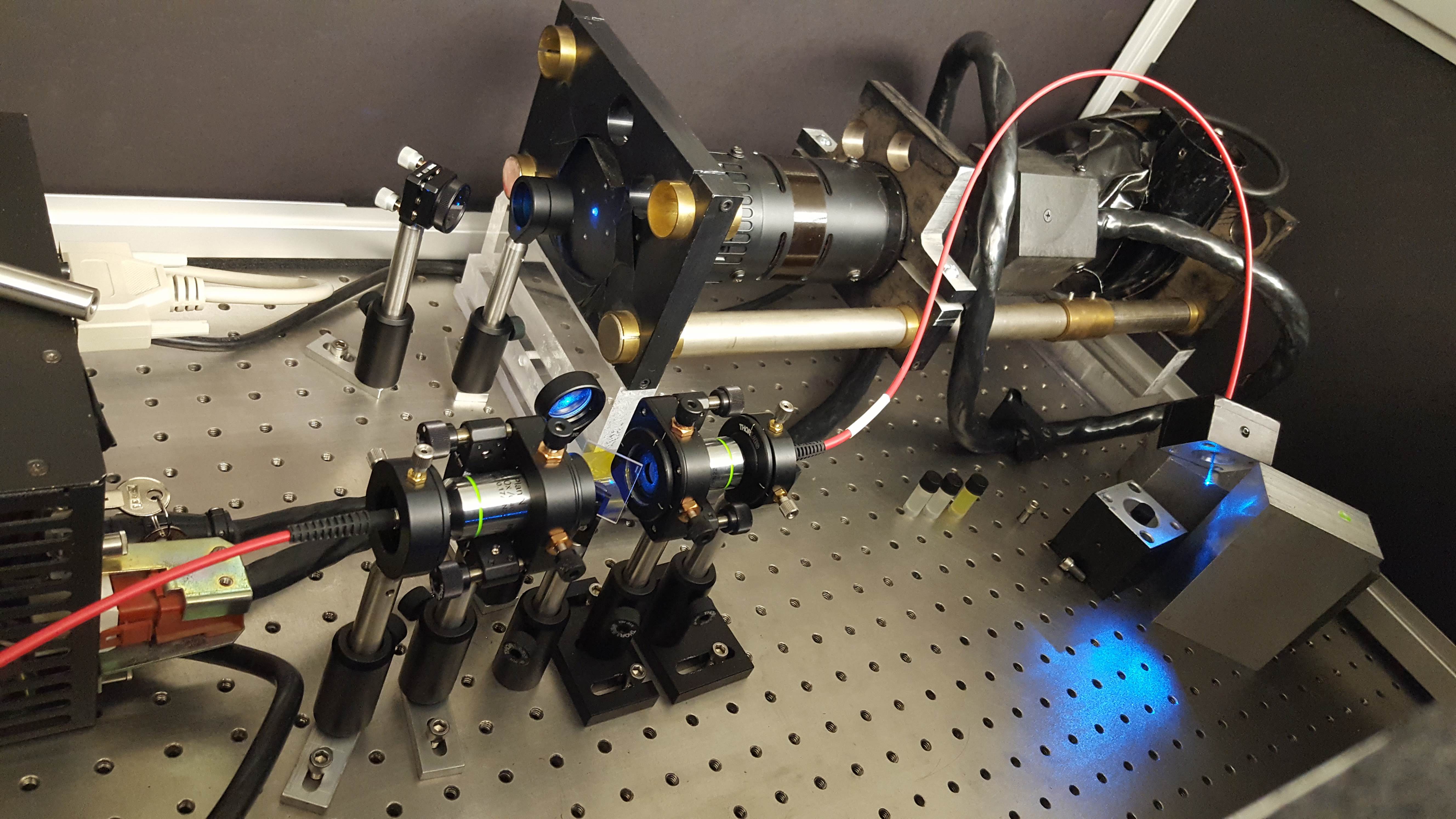}\newline \newline
    \includegraphics[width=0.60\textwidth]{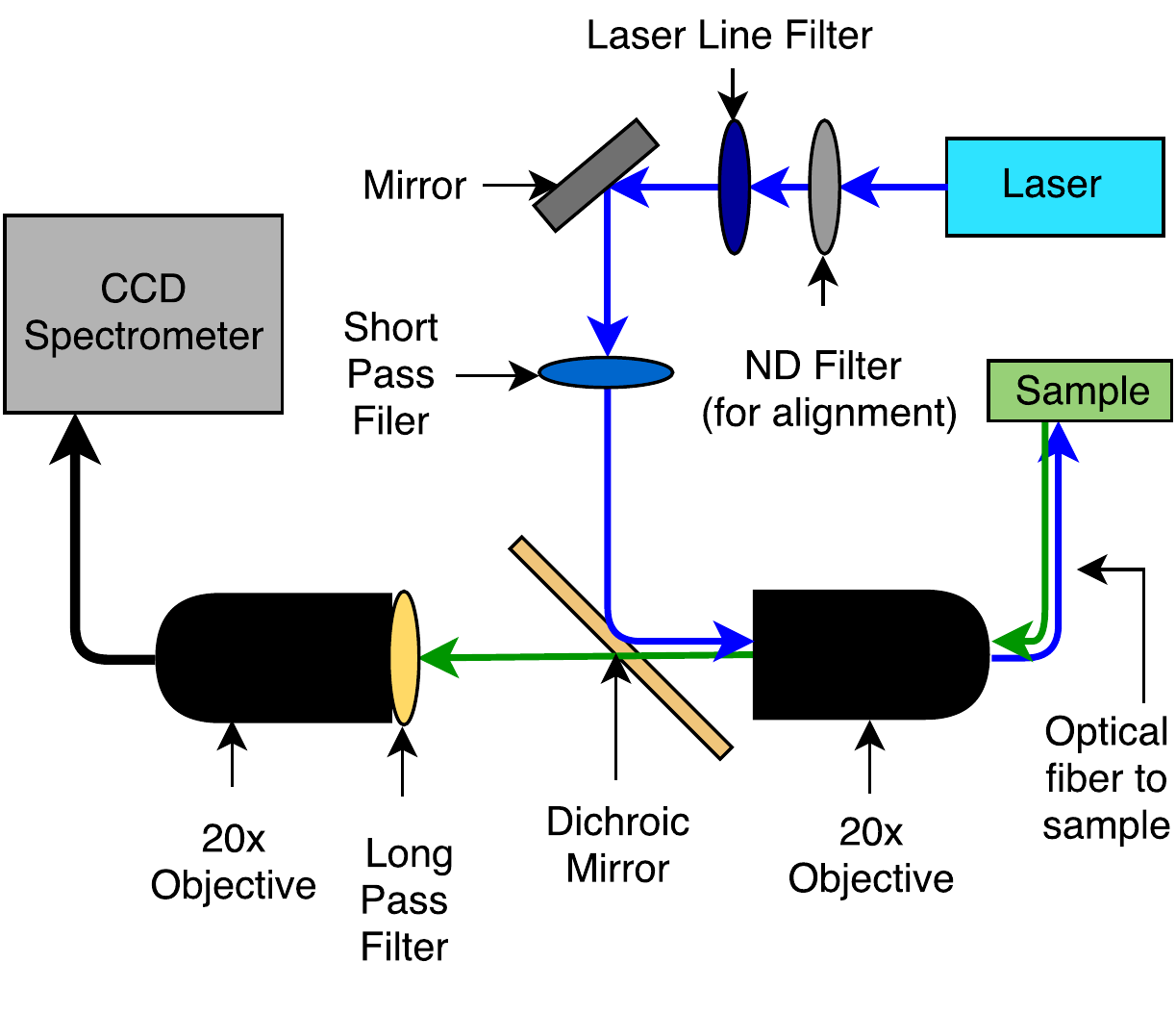}

    \caption{Photograph (top) and diagram (bottom) of the optical system.}
    \label{fig:lense}
\end{figure}

For the studies described in this paper we use a 488 nm argon ion laser as an excitation light source.  Our goal is to demonstrate sensing of barium dications at a remote location using a fiber-coupled optical system.  For gas-phase tests, this will be connected to a total internal reflection fluorescence sensor, described in Section \ref{sec:GasTests}.  The optical system is shown in Figure \ref{fig:lense}. Excitation light is coupled into a 1000~$\mu m$ high numerical aperture (0.5 NA) multimode fiber using a 20$\times$ microscope objective (Olympus pln20x with 0.4 NA).  The fiber carries the excitation light to a cuvette containing the fluorescent mixture approximately 1 m away.  To avoid introducing long-wavelength background light into the sample, the laser light is filtered first with a 488 nm laser line filter, and then with a 500~nm short pass filter, before being reflected from a dichroic mirror (DM) with 500~nm cutoff wavelength.   The excitation power was measured at the sample, and we find that in idle mode the laser produces 110~$\mu$W at sample and when run at full power can deliver 1.1~mW.

\begin{figure}[t]
    \centering
    \includegraphics[width=0.49\textwidth]{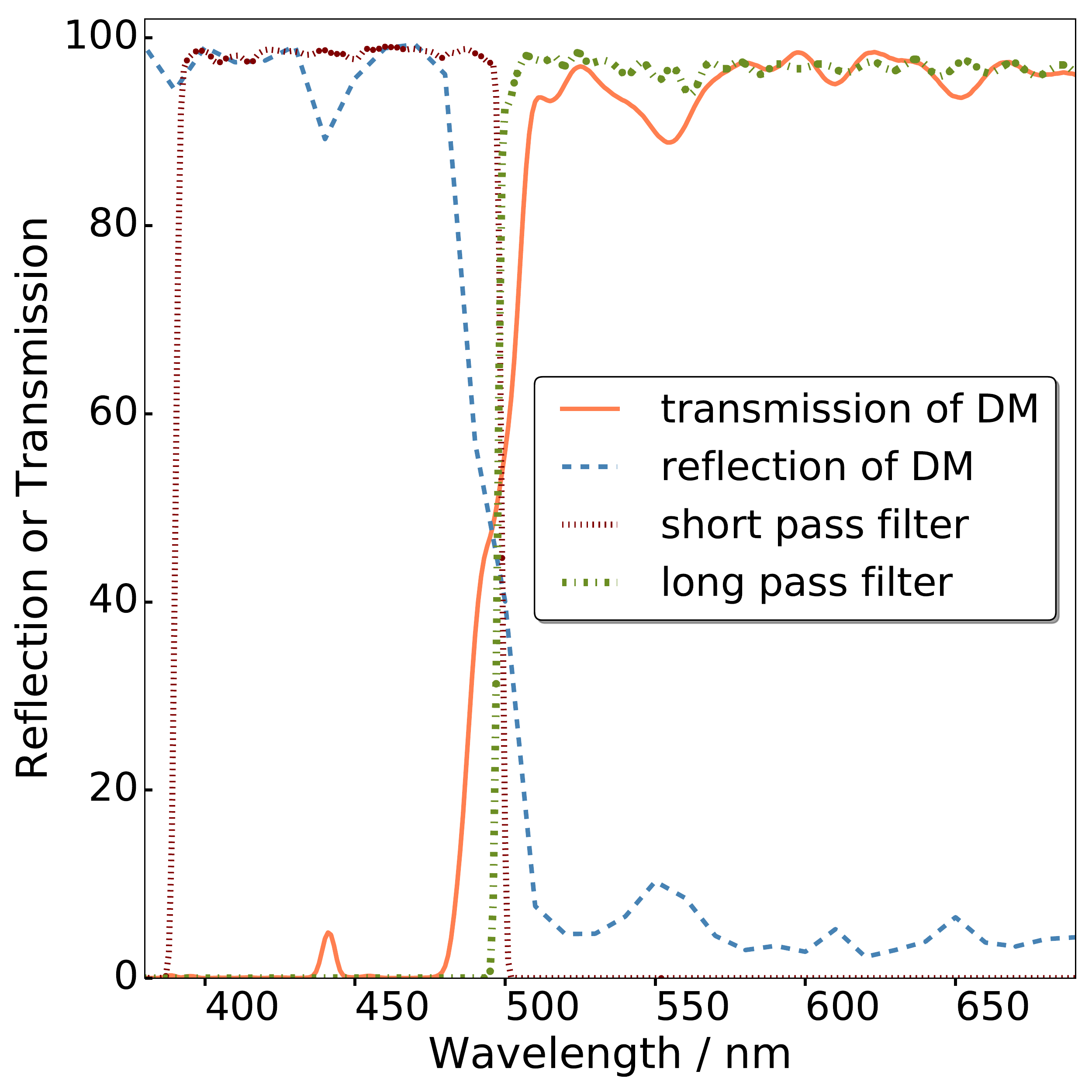}
    \includegraphics[width=0.49\textwidth]{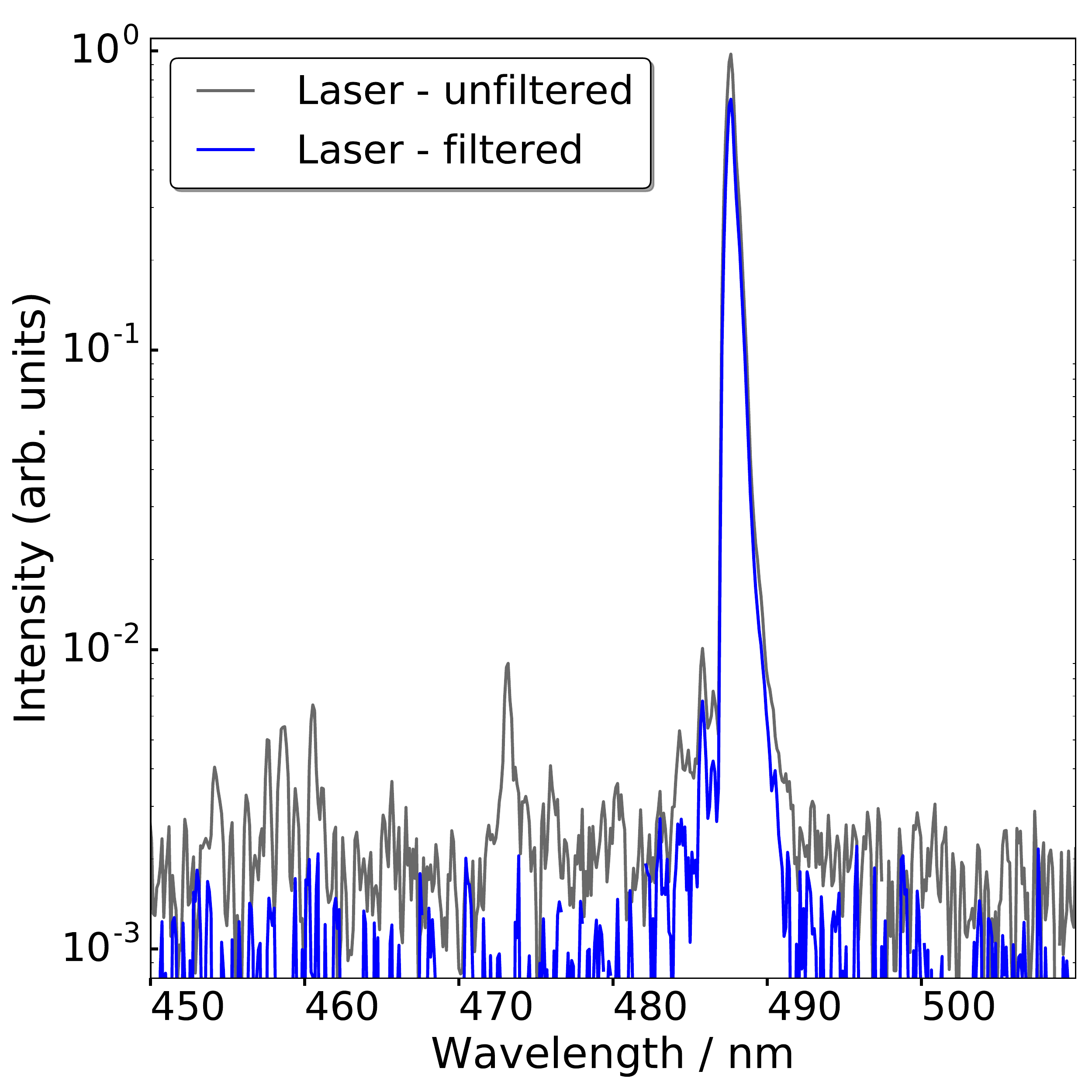}
    \caption{Left: The transmission/reflection of the DM overlayed with the transmission of the SPF and LPF. Right: Argon ion laser spectrum before and after excitation filtering.  The dominant feature is the laser peak at 488 nm. Secondary peaks such as the one at 470~nm are removed by the filter optics.}
    \label{fig:filterslaser}
\end{figure}

A fraction of the sample fluorescence emission is captured and guided back along the fiber with the >500~nm component passing through the dichroic mirror.   This long wavelength light is focussed by a second 20$\times$ objective into a fiber which runs to a CCD spectrometer (CCS100 from Thorlabs).  The excitation optics were aligned using the 488 nm laser with a neutral density filter at the aperture, and the detection optics were aligned using a red HeNe  laser focussed into the far end of the fiber to maximize transmission.  The full system is operated inside a dark box, with the spectrometer on the outside connected using an SMA optical fiber feedthrough.  A plot showing the short-pass, long-pass, and dichroic transmission and reflection spectra in Figure ~\ref{fig:filterslaser}, left, and the filtered and unfiltered laser power spectrum are shown in Figure \ref{fig:filterslaser}, right.

\begin{figure}[t]
    \centering
    \includegraphics[width=0.49\textwidth]{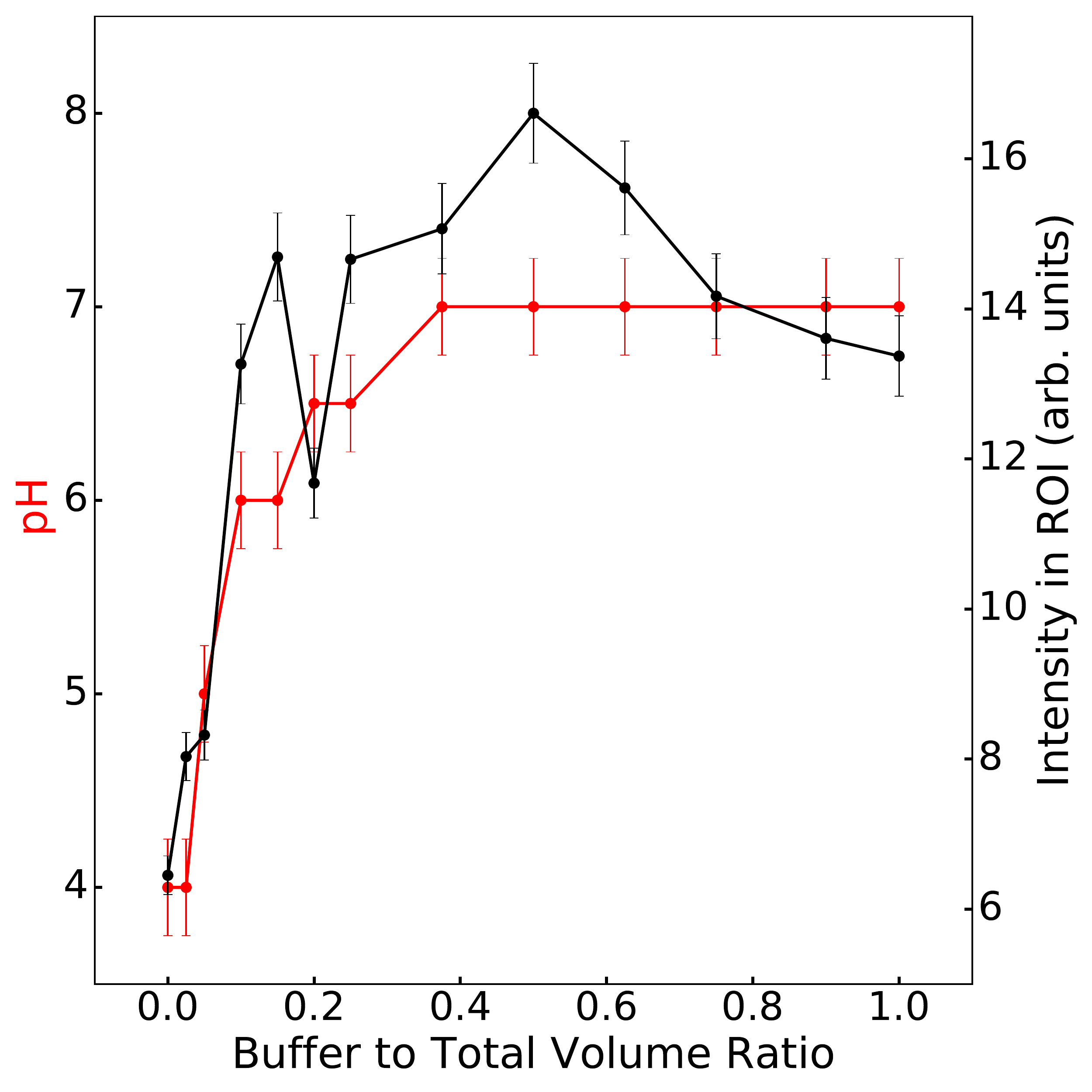}
    \includegraphics[width=0.49\textwidth]{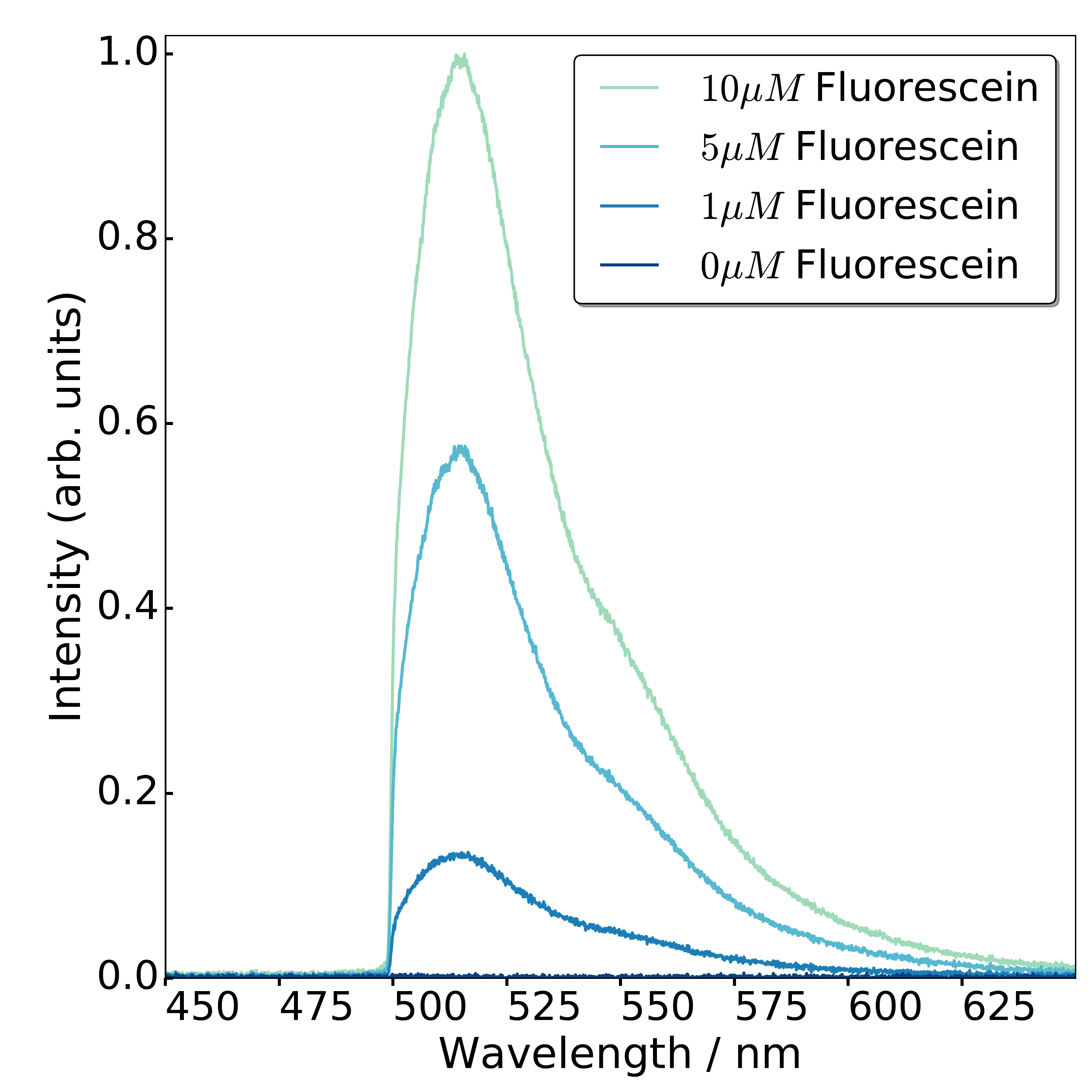}
    \caption{Left: pH dependence of fluorescein emission, which led us to stabilize solutions with imidazole / HCl buffer.  Right: detected fluoresence spectrum from pH-stabilized fluorescein solutions probed with the fiber detection system.}
    \label{fig:fluorescein_data}
\end{figure}

The samples used in these studies are aqueous solutions held in plastic cuvettes, placed inside a custom-made cuvette holder that fixes the fiber at a constant height in the solution.  The system was exercised first with fluorescein solution since it has approximately the same fluorescence spectrum as chelated Fluo-4 and can be studied independently of the degree of chelation.  In calibration runs we observed a pH dependence to the fluorescence intensity of fluorescein, shown in Figure \ref{fig:fluorescein_data}, left.  For this reason, in all subsequent studies a pH buffer consisting of imidazole and HCl is used to hold the solution at pH~7.2.  This is necessary because the ionic barium compounds we will introduce into the solution are somewhat basic, whereas Fluo-4 is a carboxylic acid derivative.  Thus, testing these at various concentrations without a buffer would introduce undesirable effects on the fluorescence from pH changes alone.  To ensure the pH was properly stabilized, several of the samples described were tested for their pH after scanning, and a consistent result pH$\sim$7 was obtained in all cases.  Such stabilization will not be required in HPGXe, since the concept of pH is not applicable in dry environments.

To make fluorescein test samples, first a 100~$\mu$M stock solution of fluorescein was mixed.
Samples made from the stock solution had a final volume of 550~$\mu$L.  They contained 50~$\mu$L of the buffer solution, and the other 500~$\mu$ L was pure water and fluorescein stock solution mixed to the desired molality by micro-pipette.  The detected fluorescence spectra for various quantities of fluorescein dye are shown in Figure \ref{fig:fluorescein_data}, right.  It is notable that following emission and excitation filtering, the background at the excitation wavelength from the laser is suppressed to negligible levels at the CCD.    An approximately linear relationship between fluorescein concentration and fluorescence intensity is observed.  The repeatability of the measurement was quantified by preparing five independent samples and scanning them, giving a spread of 2.5 \% between measurements, which is affected both by the reproducibility of the solutions and the placement of the fiber.  This number can be taken as an estimate of the systematic uncertainty on relative measurements in all subsequent studies.

Once the system was configured for sensitivity at the appropriate emission wavelengths, Ba$^{++}$ detection studies were instigated.  A stock solution of fluorophore was mixed to $100 \mu$M.  Fluo-3 and Fluo-4 are highly sensitive dyes with dissociation constant ($k_d$) in the nano-molar range, and even in the very clean water used for these studies (Sigma Aldrich ACS reagent, for ultratrace analysis) the residual calcium ion concentration of 0.2~$\mu$g/kg (5 nM) gave a significant fluorescent background.  To suppress this background we use a standard method in high-sensitivity calcium detection \cite{MolProbesBook}, which is to introduce the non-fluorescent chelator BAPTA which eliminates the majority of the free ions.  During optimization of our protocol we also found the background could be reduced by using plastic rather than glass vials and by rinsing all elements with three washes of clean water.

\begin{figure}[t]
    \centering
    \includegraphics[width=0.49\textwidth]{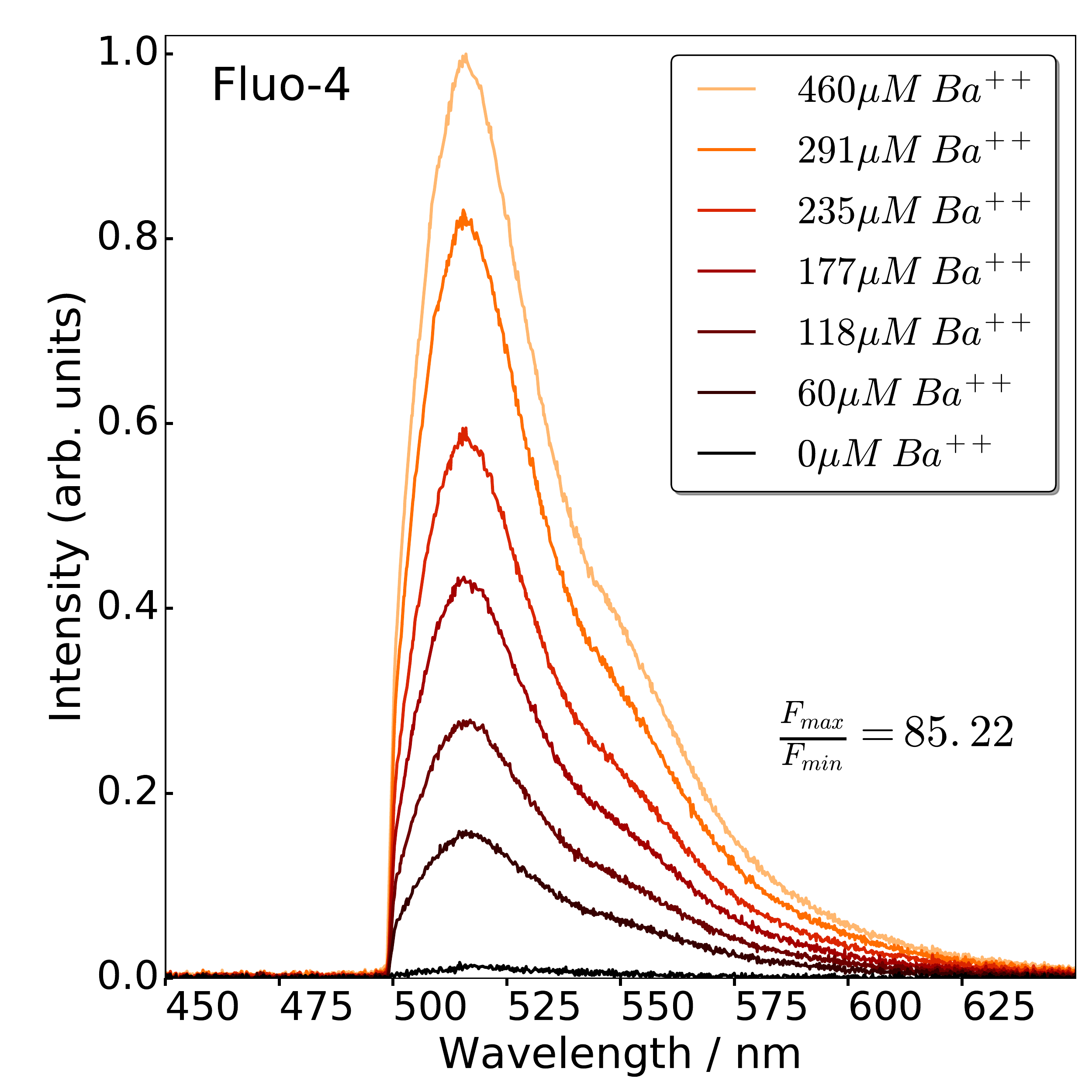}
    \includegraphics[width=0.49\textwidth]{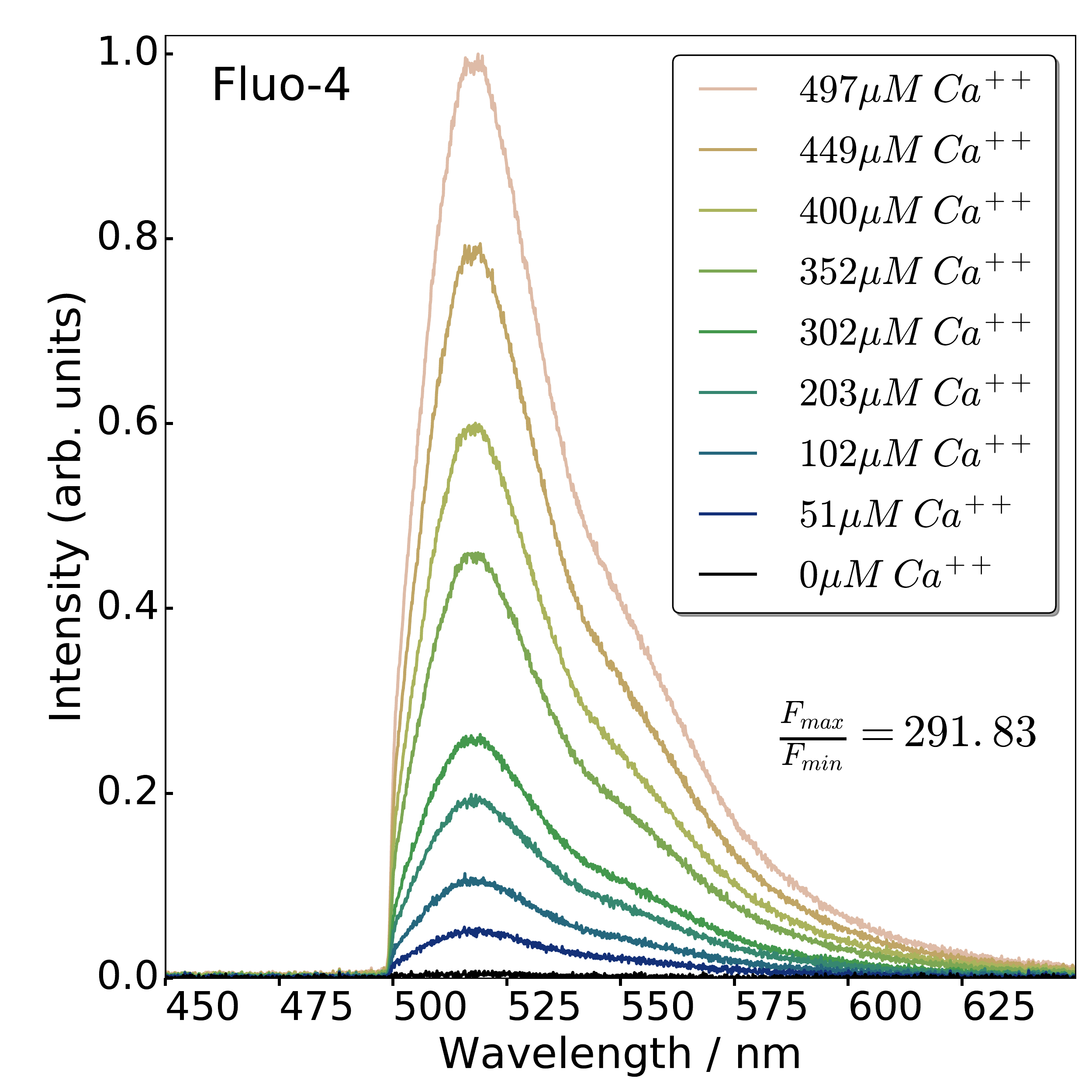}
    \includegraphics[width=0.49\textwidth]{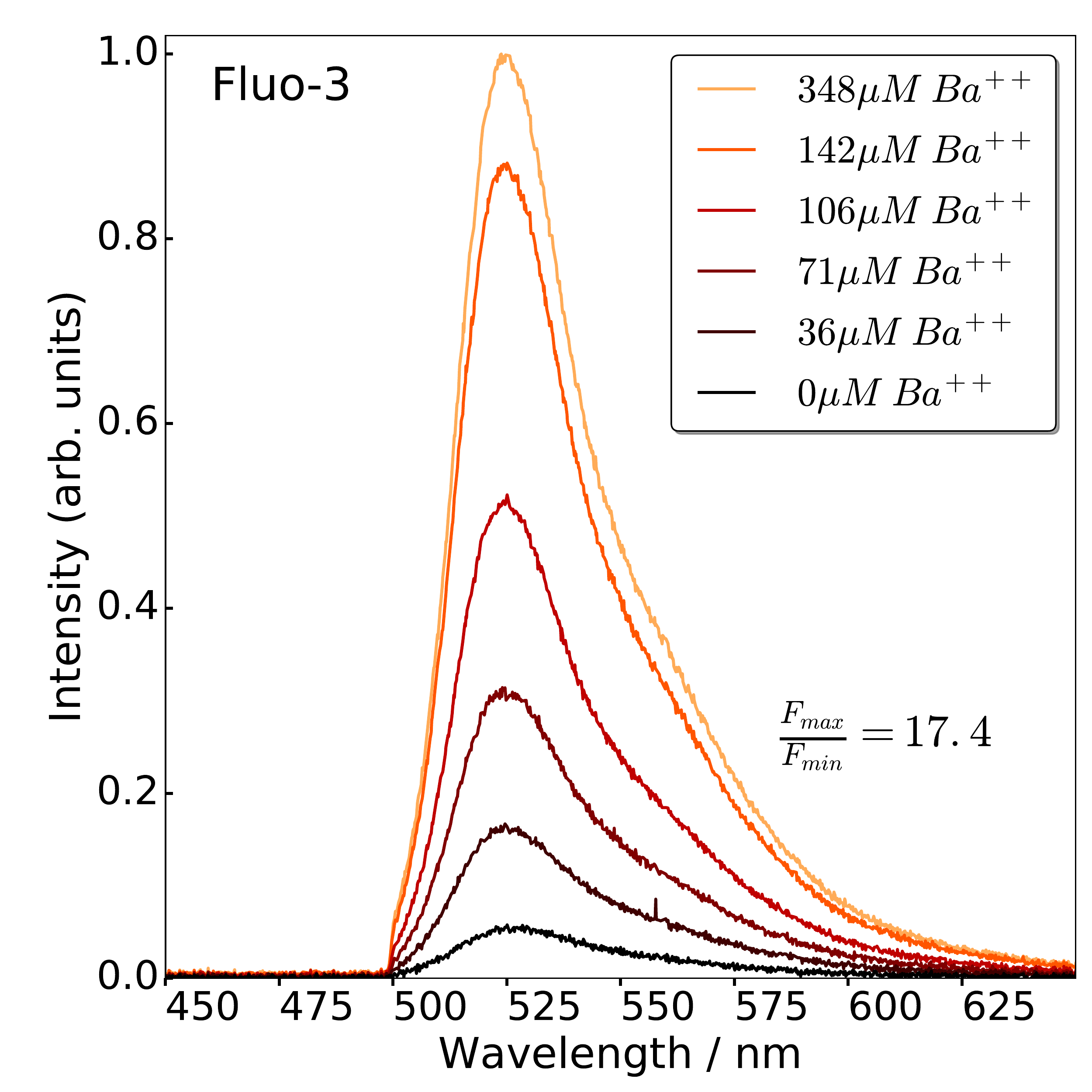}
    \includegraphics[width=0.49\textwidth]{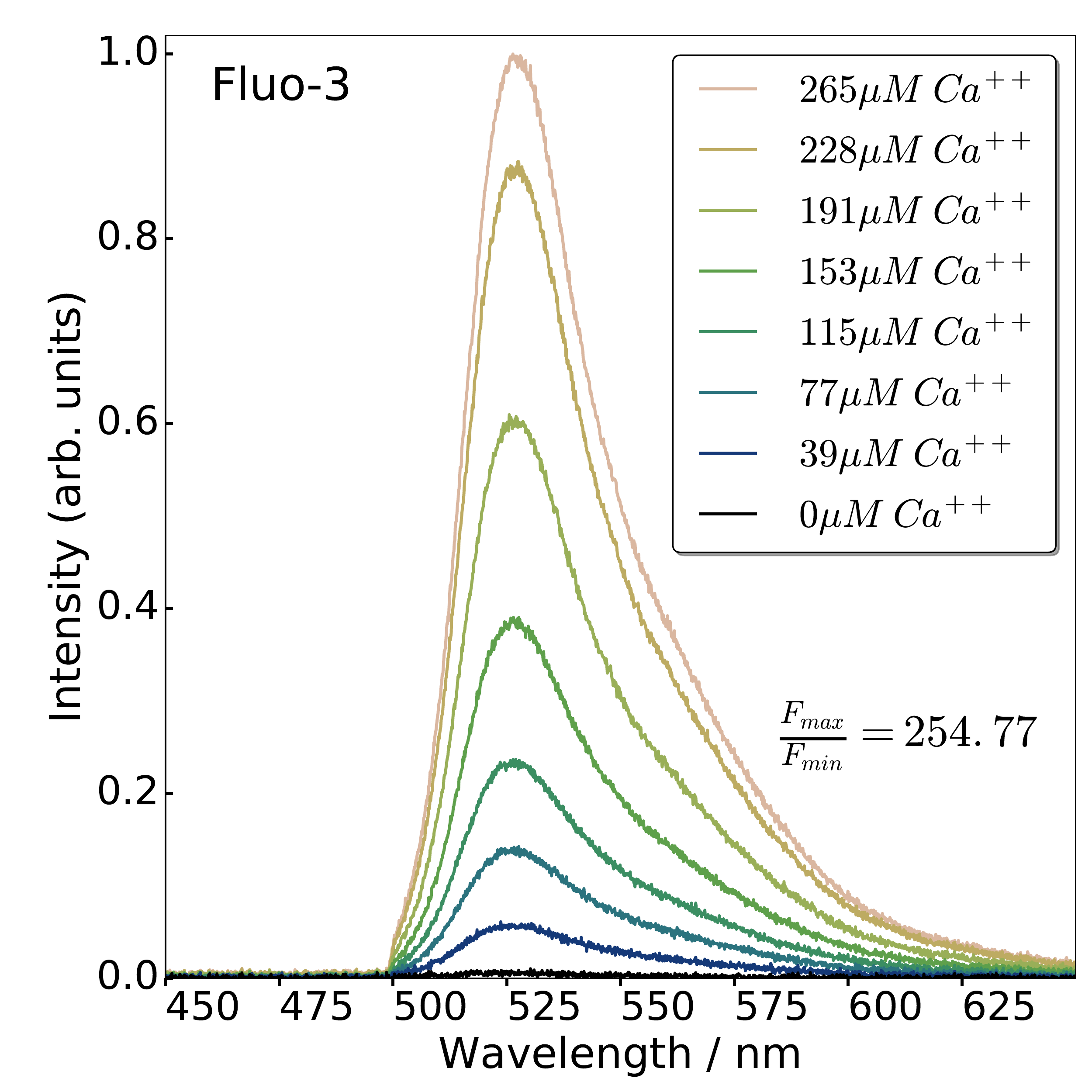}

    \caption{Ca$^{++}$ (left) and Ba$^{++}$ (right) induced fluorescence in Fluo-4 (top) and Fluo-3 (bottom) dye solutions.  These dyes, developed for calcium dication detection, show a clear affinity for barium dications, making them potentially suitable for SMFI in high pressure xenon gas TPCs.}
    \label{fig:fluo4}
\end{figure}

A sample was mixed with Fluo-3 or Fluo-4 to a volume of 500~$\mu$L,  containing 50~$\mu$L of buffer solution, 260~$\mu$M BAPTA, 10~$\mu$M of fluorophore.  We measured both the response to Ca$^{++}$ ions and the response to Ba$^{++}$ ions in different concentrations.  At some ion concentration the signal saturates at fluorescence intensity F$_{max}$, and we report the relative size of the saturated signal strength to the background fluorescence level, F$_{min}$.  Curves for intermediate ion concentrations can be normalized to these numbers.  Dications were added in the form of calcium / barium perchlorate, which dissociates readily in solution into Ca$^{++}$ / Ba$^{++}$ and ClO$_4^-$ ions.  The salt was pre-mixed at a concentration of 5.95~mM for barium perchlorate and 6.43~mM for calcium perchlorate, this concentrated solution allowing a sizable ion concentration to be added to pre-mixed samples without significant volume change or dilution.

Both calcium and barium induced fluorescence were clearly observed in both Fluo-4 and Fluo-3 samples, at concentrations between 30$\mu M$ and 500~$\mu$M.  At the peak fluorescence intensity, the signal to background ratio (F$_{max}$/F$_{min}$) is 85.22 for barium and 291.83 for calcium in Fluo-4, and 17.40 for barium and 254.77 for calcium in Fluo-3.  The background before any dications were added is believed to derive from uncaptured free Ca$^{++}$ and other free metal ions in the purified water.  The maximum signal to background ratio is driven by this background intensity and by the saturation point of the dye with each cation type.  Both dyes show a higher affinity for calcium than barium, but the difference in affinities is significantly smaller for Fluo-4 than Fluo-3, as shown by the improved signal-to-background ratio.  The emission spectrum for barium- and calcium-chelated dyes are found to be almost identical in shape, though both Fluo-3 and Fluo-4 have a barium-chelated spectrum peaking at wavelengths 2-3 nm shorter than the calcium-chelated spectrum.

The clear observation of barium-induced fluorescence demonstrates that 1) Fluo-3 and Fluo-4 are suitable dyes for barium sensing and that 2) our remote scanning system is capable of barium ion detection at the end of a fiber.  The small residual background, though not problematic for this investigation, will be the focus of further optimization.  Although no free Ca$^{++}$ or other doubly charged metal dications are expected in HPGXe detectors, preparation of barium-sensitive coatings which are free of pre-chelated molecules is an important requirement for production of sensitive single-barium-ion sensors.  Other chelators including EGTA and EDTA \cite{McGuigan1991}, as well as calcium sponges with BAPTA-like molecules bonded onto polystyrene beads \cite{ThermoCaSpong} are available for these purposes and are being investigated. Furthermore, as well as suppressing the free-ion background, BAPTA also competes with the SMFI dyes when dications are introduced, thus reducing fluorescence response per ion.  For this reason, addition of BAPTA in sensors for dry environments is to be avoided where single ion sensitivity is desired.

\section{Prospects for testing SMFI barium sensors in xenon gas}
 \label{sec:GasTests}

In the previous section we described the development of a system for sensing production of barium dications using a fiber-coupled optical system.  The next stage of this work is to translate these studies from an aqueous environment to a dry noble gas.  We are presently constructing a high-pressure xenon test stand with a barium plasma source which will deliver a beam of Ba$^{++}$, separated from the plasma by time-of-flight, into a sensing region where barium-sensitive electrodes can be tested.  This device will be used to establish whether barium-sensitive SMFI dyes continue to function in HPGXe environments.

\begin{figure}[b]
\begin{centering}
\includegraphics[width=0.99\columnwidth]{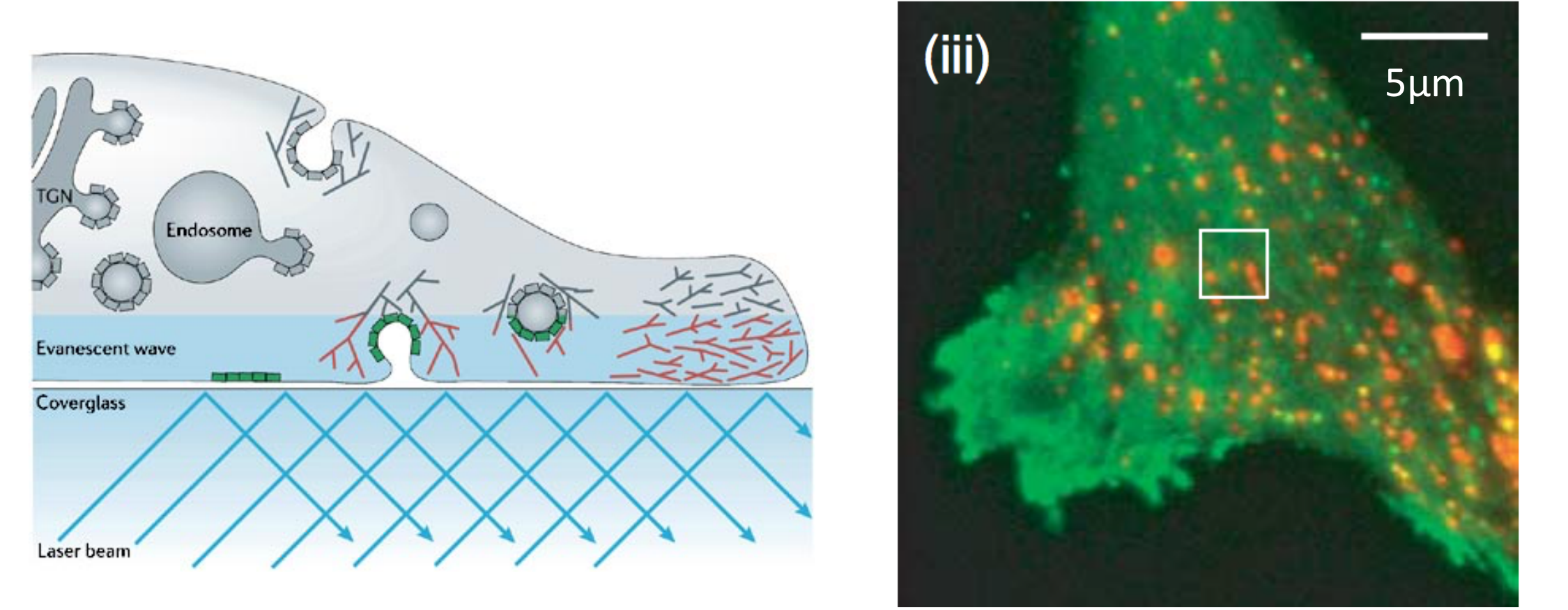}
\par\end{centering}

\caption{Left: illustration showing the use of TIRF to identify actin polymerization bursts in mammalian cells, from  \cite{Kaksonen2006}.  Right: image from a similar study, published in \cite{Merrifield2015}   \label{fig:TIRFExample}}
\end{figure}

The barium sensing concept we will deploy is based on total internal reflection fluorescence (TIRF) microscopy.  In TIRF, the evanescent electromagnetic wave at a light-guide or fiber surface excites fluorophores within a few hundred nanometers of the interface.  This is shown schematically in Figure \ref{fig:TIRFExample}, left from \cite{Kaksonen2006}, and is a common microscopy technique in biological sciences.  It is especially appropriate for tasks where specific sensitivity to near-surface fluorophores is desirable.  Figure \ref{fig:TIRFExample}, right shows data from \cite{Merrifield2015}   where TIRF was used to reveal the actin polymerization bursts at endocytic sites in mammalian cells.  Reviews of the technique applied to both single- and multiple-molecular fluorescence imaging can be found in  \cite{Fish2009, stuurman2006imaging}.

For our detection elements, we will place a relatively negative cathode behind a coated fiber to focus drifting positive ions onto a region where a coating of barium-sensitive dye has been deposited onto the fiber surface.  Because the fluorescent molecules of the coating are emitting in the near-field regime relative to the dielectric interface, evanescent excitation of the fluorophore will result in some fluorescence emission back into the fiber.  This returning light will be dichroically separated, using the same system used in the aqueous studies of Section \ref{sec:LiquidTests}.  

The major questions to be addressed by these studies are:

\begin{figure}[t]
\begin{centering}
\includegraphics[width=0.99\columnwidth]{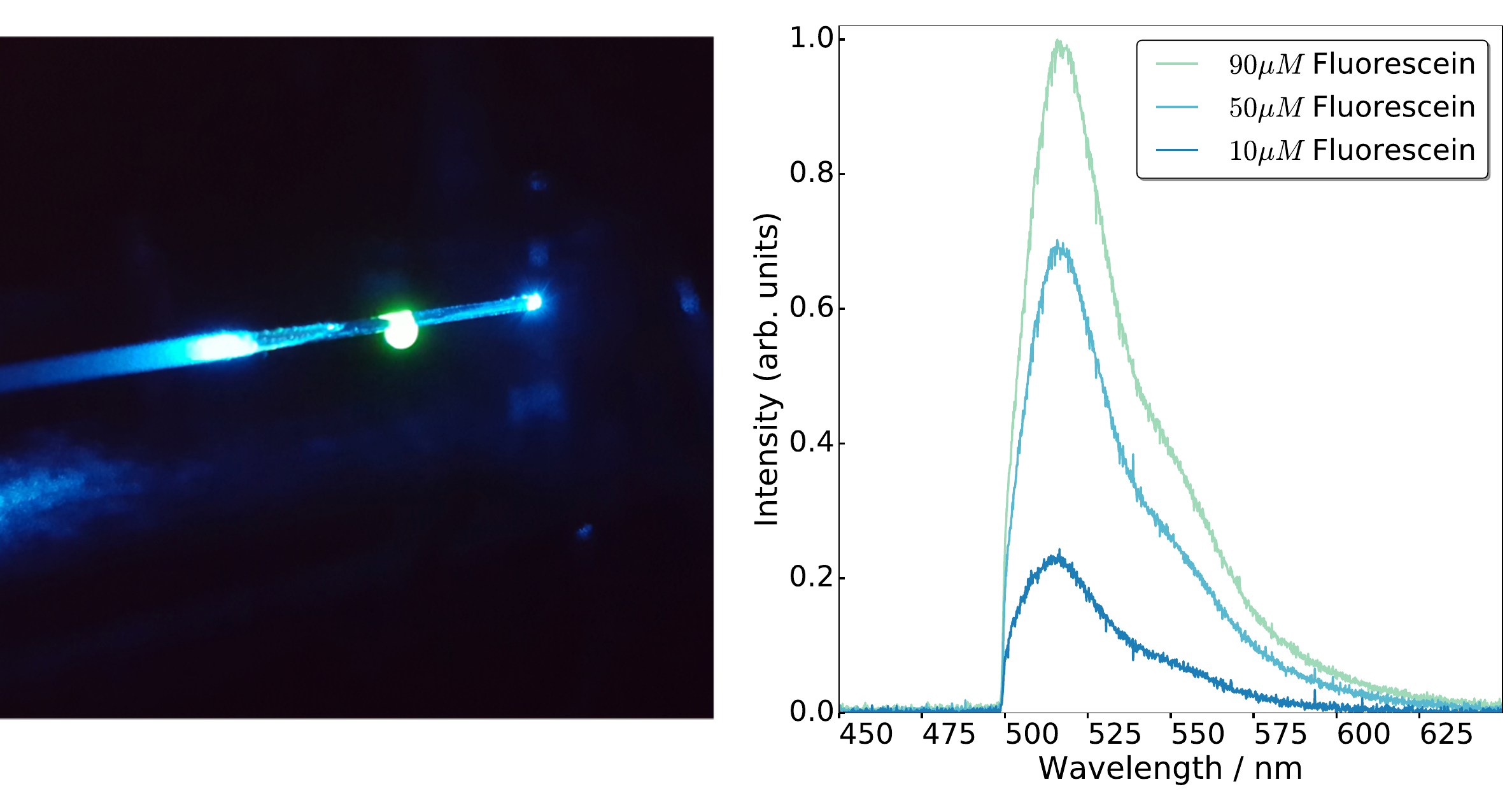}
\par\end{centering}

\caption{Left: Photograph showing TIRF excitation. Visible green fluorescence light emanates from a bead of liquid at the side of a fiber.  Right: Fluoresence response as seen at the spectrometer, showing that some of the emitted light couples back into the fiber modes due to TIRF emission in the near-field regime. \label{fig:TIRFLabDemo}}
\end{figure}

\begin{itemize}
  \item Can SMFI dyes be used for ion detection in a non-aqueous environment?
  \item What are the fluorescence yields and spectroscopic properties of such dyes in HPGXe?
  \item What is the barium capture efficiency in HPGXe?
  \item What are the drift properties (mobility, diffusion, etc) of Ba$^{++}$ in HPGXe and can it be efficiently focused onto a detection element?
\end{itemize}

Our TIRF-based sensors can only be properly operated in a dry environment, since the coating is water-soluble.  However, as a first test of our sensitivity to TIRF fluorescence a preliminary study was made using a 4~$\mu$L droplet of fluorescein solution.  A clean fiber with the outer jacket removed was placed with its side in the droplet.  Fluorescence light is generated which is both visible by eye and detectable in our spectrometer.  Light emitted outside the fiber cannot enter a fiber mode unless it was produced by a fluorophore in the near-field regime to the fiber/water interface - this tells us that the detected green light seen at the spectrometer is a result of TIRF fluorescence.  A photograph of this test and the spectrometer output are shown in Figure \ref{fig:TIRFLabDemo}. Using this system and a barium sensitive SMFI dye coating, we plan to use the methods described in this paper to detect Ba$^{++}$ ions isolated by our HPGXe barium ion source.  This work will be presented in future publications.

\section{Conclusions}

To discover Majorana neutrinos even in the most pessimistic cases allowed by the inverted mass ordering requires ton-scale experiments and a background rejection capability sufficient to achieve $b$ < 0.1 ct (ROI) ton$^{-1}$ yr$^{-1}$.  Meeting this challenge requires suppression of backgrounds rates by a factor of 40 to 3000, perhaps beyond the capabilities of existing technologies. Thus, the technical challenge presented to the $0\nu\beta\beta$ field is very substantial, if not daunting.

Techniques that can reliably identify the daughter ion in $0\nu\beta\beta$ experiments are strongly motivated since these could provide a nearly background-free positive criterion for discovery. In this paper we have outlined a new concept for barium daughter tagging in high-pressure xenon gas TPCs using the technique of SMFI. SMFI techniques are routinely used with single molecule sensitivity in living cells and aqueous media and may hold promise for ton-scale $0\nu\beta\beta$ detectors.

By using a dye optimized for sensitivity to Ca$^{++}$ we have developed a sensor which can detect the presence of bulk Ba$^{++}$ ions in aqueous solution using a fiber-coupled optical system.  Signal to background ratios as large as 85 have been achieved for barium detection with this device.  This sensitivity is sufficient for explorations of whether this dye and others like it can capture barium ions and exhibit Ba$^{++}$-induced fluorescence in dry noble environments.  

The next step of this R\&D program involves operating a total internal reflection fluorescence (TIRF) sensitive electrode coupled to our optical system under exposure to a pure Ba$^{++}$ beam in xenon gas.   Positive results would suggest significant promise in the SMFI barium tagging concept and a possible path to discovery for the field of neutrinoless double beta decay.

\acknowledgments

We thank Rasika Dias and Sandy Dasgupta of UT Arlington for helpful chemistry guidance, and Sebastian Raquena and Zygmunt (Karol) Gryczynski of Texas Christian University for valuable early discussions about SMFI.  Thanks also to Jonathan Asaadi for a careful reading of this paper and insightful comments.  This work was supported by the University of Texas at Arlington.

\section*{References}

\printbibliography[heading=none,maxnames=99]

\end{document}